\documentclass[12pt]{article}
\def\timesbox{\hbox{$\scriptscriptstyle\times$}}
\def\ant{ {{\lower 1ex  \timesbox} \atop {\raise 1.5ex  \timesbox}}}
\newcommand{\Zop}{{\hbox{ Z\kern-1.6mm Z}}}\begin{scriptsize}\end{scriptsize}
\newcommand{\beq}{\begin{equation}}
\newcommand{\eeq}{\end{equation}}
\newcommand{\bea}{\begin{eqnarray}}
\newcommand{\eea}{\end{eqnarray}}
\newcommand{\ra}{\rangle}
\newcommand{\la}{\langle}
\newcommand{\lt}{\left}
\newcommand{\rt}{\right}
\newcommand{\Iop}{\relax{\rm I\kern-.18em I}}
\newcommand{\one}{{\hbox{ 1\kern-1.2mm l}}}
\newcommand{\T}{{\cal T}}

\newcommand{\dt}{\delta}
\newcommand{\del}{\partial}

\newcommand{\eps}{\epsilon}

\newcommand{\lam}{\lambda}

\newcommand{\s}{\sigma}

\newcommand{\xh}{\hat x}
\newcommand{\ph}{\hat p}
\newcommand{\pih}{\hat \pi}

\begin{document}

{}~
{}~
\break

\vskip 2cm

\centerline{\Large \bf DeWitt-Virasoro construction}

\medskip

\vspace*{4.0ex}

\centerline{\large \rm Partha Mukhopadhyay }

\vspace*{4.0ex}

\centerline{\large \it The Institute of Mathematical Sciences}
\centerline{\large \it C.I.T. Campus, Taramani}
\centerline{\large \it Chennai 600113, India}

\medskip

\centerline{E-mail: parthamu@imsc.res.in}

\vspace*{5.0ex}

\centerline{\bf Abstract}
\bigskip

We study a particular approach for analyzing worldsheet conformal
invariance for bosonic string propagating in a curved background using
hamiltonian formalism. We work in the Schr$\ddot{\rm o}$dinger picture
of a single particle description of the problem where the particle
moves in an infinite-dimensional space. Background independence is
maintained in this approach by adopting DeWitt's
(Phys.Rev.85:653-661,1952) coordinate independent formulation of
quantum mechanics. This enables us to construct certain  background
independent notion of Virasoro generators, called DeWitt-Virasoro
(DWV) generators, and invariant matrix elements of an arbitrary
operator constructed out of them in spin-zero representation. We show
that the DWV algebra is given by the Witt algebra with additional
anomalous terms that vanish for Ricci-flat backgrounds. The actual
quantum Virasoro generators should be obtained by first introducing
the vacuum state and then normal ordering the DWV generators with
respect to that. We demonstrate the procedure in the simple cases of
flat and pp-wave backgrounds. This is a shorter version of
arXiv:0912.3987 [hep-th] with many technical derivations omitted.

\begin{flushleft}
\end{flushleft}

\newpage

\tableofcontents

\baselineskip=18pt

\section{Introduction and summary}
\label{s:intro}

Equations of motion (EOM) for backgrounds in string theory are derived
from the condition of worldsheet conformal invariance\footnote{Some of
  the original references are \cite{friedan80, lovelace84, fradkin85,
    sen85, callan85, fridling86}.}. This has mostly been studied by
computing the beta functions of the nonlinear sigma model using
background field method \cite{friedan80}. A BRST hamiltonian approach
was also discussed to some extent in the literature \cite{maharana87,
  akhoury87, das87, fubini89, alam89, diakonou90, wess90} (see also
\cite{jain87}) where one attempts to verify the constraint algebra at
the quantum level\footnote{Recently a similar method has been used
  \cite{kazama08, pm08, pm0902, pm0907} to study exact conformal
  invariance of the worldsheet theory in type IIB R-R plane-wave
  background \cite{blau01}.\label{iib}}. 

In this work we explore a new approach for studying the same problem
in hamiltonian framework. In this approach one attempts to find a
suitable string generalization of DeWitt's coordinate independent
description of particle quantum mechanics as discussed in
\cite{dewitt52, dewitt57}\footnote{DeWitt's formulation was applied to
  string theory earlier in \cite{lawrence95}.}. We attempt to develop
the framework in analogy with ordinary particle quantum mechanics
where, given a classical hamiltonian, the quantum mechanical problem
can be solved in two steps, namely (1) find the quantum hamiltonian
using which the Schr$\ddot{\rm o}$dinger equation needs to be written
down, (2) find the vacuum (and other excited states) by solving the
Schr$\ddot{\rm o}$dinger equation. In case the second step can not be
carried out exactly, one develops a perturbative method for solving
the problem approximately. The first step is enough to provide the
complete quantum definition for a single particle system. However, for
a quantum field theoretic system, as the case at hand, it serves the
purpose only in a limited sense because of the divergences arising
from the infinite number of degrees of freedom. Such divergences need
to be cured by introducing the vacuum state. 
In this work we will address the question analogous to the first step
only leaving the general study of the vacuum for future work. However,
we will discuss the latter in the specific cases of flat and pp-wave
backgrounds. Finding out how to follow through the second step in
general is an interesting question that needs further
investigation. One may also wonder at this point how even the first
step itself can be carried out in presence of divergences. As we will
see, manifest general covariance will enable us to manipulate various
expressions containing such divergences. 

We will now describe our construction in detail. This is done by first
developing a single particle description of the worldsheet theory. At
the classical level such a description is obtained by re-writing the
theory in terms of suitable Fourier modes such that it describes a
particle moving in an infinite-dimensional curved space (subject to
certain potential)\footnote{This way the infinite number of degrees of
  freedom of the string is given an interpretation of number of
  spacetime dimensions.}. This is the kind of non-linear system that
was studied in \cite{dewitt52, dewitt57}. The general coordinate
transformation (GCT) in the infinite-dimensional spacetime (induced by
the same in the physical spacetime where the string is moving) is
interpreted as a point canonical transformation in the single particle
classical mechanical problem. The latter can then be identified as a
subgroup of all unitary transformations in the corresponding quantum
theory. DeWitt's analysis shows how the quantum theory can be written
down in a manifestly covariant manner in position representation. 

There are a few generalizations involved in our work from DeWitt's original work. The analysis of
\cite{dewitt52} considered a non-relativistic particle so that the
general covariance was sought only for the spatial slice. In our case
we adopt the infinite-dimensional language for the matter part of the
worldsheet theory in conformal gauge. The resulting particle-like
theory looks like a worldline theory with full covariance in
spacetime. A more important difference is having infinite number of
Virasoro generators instead of only the hamiltonian as in DeWitt's
case. Because of the presence of infinite number of dimensions the
theory possesses certain {\it shift} properties which look unusual
from the particle point of view. These properties dictate the behavior
of the theory under certain shift of the spacetime dimensions,
i.e. the string modes. Since Virasoro generators relate different
string modes, such shift properties are inherently related to the
existence of these generators. 

The above generalization enables us to construct a background
independent quantum\footnote{We will use the word {\it quantum} in the
  limited sense of step (1) above except for the discussion in section
  \ref{s:special} where vacuum and normal ordering are discussed for
  flat and pp-wave backgrounds.} version of the Virasoro generators,
hereafter called DeWitt-Virasoro (DWV) generators, and coordinate
invariant matrix elements involving them between two arbitrary scalar
states. As expected, all such expressions contain divergences. But
such divergences are hidden in the form of infinite-dimensional traces
and therefore our expressions can be manipulated in the formal
sense. We then go ahead and compute the algebra of DWV generators in
spin-zero representation. Interestingly, we find that the result is
given by the Witt algebra with additional operator anomaly terms that
vanish for Ricci-flat backgrounds. A few comments are in order, 
\begin{itemize}
\item
The central charge term is expected to appear once the vacuum is
introduced and the DWV generators are properly normal ordered. We
demonstrate this for flat and pp-wave backgrounds. 

\item
The Ricci-anomaly term is a quantum contribution at the leading order
in $\hbar=\alpha'$. Such a term arises because of our covariant
treatment. This is an interesting result as Ricci-flatness is also
found to be the condition for one-loop conformal invariance in the
lagrangian framework \cite{friedan80}. However, since a perturbative
expansion has not yet been formulated in our work, {\it a
  priori} it is not clear how to relate the two results in a precise way. 

\item
As mentioned in footnote \ref{iib}, exact conformal invariance of the
worldsheet theory in type IIB R-R plane-wave background was studied
using the hamiltonian method in \cite{kazama08, pm08, pm0902,
  pm0907}. However, general covariance was not made manifest in these
works. It was pointed out in \cite{pm0902} that this non-covariant
computation leads to certain Virasoro anomaly terms that suffer from
operator ordering ambiguity. 

Here we show that considering the pp-wave as a special case of the
present covariant framework improves the understanding of how the
above ambiguity may be fixed. 
\end{itemize} 

The rest of the paper is organized as follows. The
infinite-dimensional language is explained in
sec.\ref{s:map}. Construction of the DWV generators has been discussed
in sec.\ref{s:vir-gen}. We summarize the results for the DWV algebra
in sec.\ref{s:vir-alg}. We discuss the flat and the pp-wave
backgrounds as special cases of the present construction in
sec.\ref{s:special}.  

This article is a shorter version of the work in \cite{pm0912}. Many
of the technical derivations, which are omitted here, can be found in
that work. 

\section{Mapping to infinite dimensions}
\label{s:map}

We consider a bosonic closed string propagating in a $D$ dimensional curved background, hereafter
called the physical spacetime, with metric $G_{\mu \nu}$. We work in the
conformal gauge of the worldsheet theory so that the ghosts are given by the standard $(b,c)$
systems. For the purpose of the present work we will be concerned only with the matter part of
the theory. The relevant classical lagrangian is given by,
\bea
L &=& {1\over 2} \oint {d\s \over 2\pi} ~G_{\mu \nu}(X(\s))\lt[\dot X^{\mu}(\s) \dot X^{\nu}(\s)
- \del X^{\mu}(\s) \del X^{\nu}(\s)\rt]~,
\label{Lworldsheet}
\eea
where $\oint \equiv \int_0^{2\pi}$, $\mu = 0,1, \cdots , D-1$. A dot and a $\del$ denote
derivatives with respect to worldsheet time-coordinate $\tau$ and space-coordinate $\s$
respectively.
We recast this lagrangian in a form that describes a single particle moving in an
infinite-dimensional curved spacetime subject to certain potential,
\bea
L(x, \dot x) &=& {1\over 2} g_{ij}(x) \lt[\dot x^i \dot x^j -  a^i(x) a^j(x) \rt]~,
\label{Linfinite}
\eea
where $x^i$ are the general coordinates of the infinite-dimensional
spacetime. The index $i$ is given by an ordered pair of indices,
\bea
i = \{\mu, m \}~,
\eea
where $m \in Z$ is the string-mode-number such that\footnote{Throughout
the paper we will make the following type of index identifications: $i=\{\mu,
m \}$, $j=\{\nu, n \}$, $k=\{\kappa, q\}$. \label{index}},
\bea
x^i &=& \oint {d\s \over 2\pi}~ X^{\mu}(\s) e^{-im\s} ~,\cr
g_{ij}(x) &=& \oint {d\s \over 2\pi} ~G_{\mu \nu}(X(\s))e^{i(m+n)\s}~, \cr
a^i(x) &=&  \oint {d\s \over 2\pi}~ \del X^{\mu}(\s) e^{-im\s}~.
\label{xga-XGdX}
\eea

We will mainly work using the infinite-dimensional language. Below we discuss certain
properties of this language that will be relevant for our study.
\begin{enumerate}
\item
We have claimed that the worldsheet theory (\ref{Lworldsheet}) has an interpretation to be generally covariant in the infinite-dimensional sense. To see this explicitly let us consider a GCT in the physical spacetime: $X^{\mu}\to X'^{\mu}(X)$ with Jacobian matrix $\Lambda^{\mu}_{~\nu}(X) = {\del X'^{\mu} \over \del X^{\nu}}$ and its inverse
$\Lambda_{\mu}^{~\nu}(X) = {\del X^{\nu} \over \del X'^{\mu}}$. This induces a GCT in the
infinite-dimensional spacetime: $x^i \to x'^i$ such that the Jacobian matrix
$\lambda^i_{~j}(x)={\del x'^i \over \del x^j}$ and its inverse $\lambda_i^{~j}(x)={\del x^j\over
\del x'^i}$ are given by,
\bea
\lambda^i_{~j}(x) = \oint {d\s \over 2\pi}~ \Lambda^{\mu}_{~\nu}(X(\s)) e^{i(n-m)\s}~, \quad
\lambda_i^{~j}(x) = \oint {d\s \over 2\pi}~ \Lambda_{\mu}^{~\nu}(X(\s)) e^{i(m-n)\s}~.
\eea
One can then show (see \cite{pm0912}) that $g_{ij}(x)$ and $a^i(x)$ transform as tensors,
\bea
g'_{ij}(x') = \lambda_i^{~k}(x) \lambda_j^{~k'}(x) g_{kk'}(x)~, \quad
a'^i(x') = \lambda^i_{~j}(x) a^j(x)~.
\label{GCT-g}
\eea

\item
Using the map in (\ref{xga-XGdX}) one can relate any field in the infinite-dimensional spacetime
constructed out of the metric, its inverse, $a^i(x)$ and their derivatives to a non-local
worldsheet operator. A class of examples, which will prove to be useful for us, is given by a
multi-indexed object $u^{i_1j_1\cdots}_{i_2j_2\cdots}(x)$ constructed out of the metric, its
inverse, their derivatives and $a^i(x)$ (but not its derivatives) such that
$u^{i_1j_1\cdots}_{i_2j_2\cdots}(x)$ can not be factored into pieces which are not contracted
with each other. In this case one can construct a local worldsheet operator
 $U^{\mu_1 \nu_1 \cdots}_{\mu_2 \nu_2 \cdots}(X(\s))$ simply
by performing the following replacements in the expression of
$u^{i_1j_1\cdots}_{i_2j_2\cdots}(x)$,
\bea
g_{ij}(x) \to G_{\mu \nu}(X(\s))~, \quad g^{ij}(x) \to G^{\mu \nu}(X(\s))~,\quad \del_i \to
\del_{\mu}~, \quad a^i(x) \to \del X^{\mu}(\s)~.
\label{replace}
\eea
As was shown in \cite{pm0912}, the two objects $u^{i_1j_1\cdots}_{i_2j_2\cdots}(x)$ and $U^{\mu_1 \nu_1 \cdots}_{\mu_2 \nu_2 \cdots}(X(\s))$ are related to each other by the following general rule,
\bea
u^{i_1j_1\cdots}_{i_2j_2\cdots}(x) \sim [2\pi \delta (0)]^N \oint {d\s \over 2\pi} ~U^{\mu_1
\nu_1 \cdots}_{\mu_2 \nu_2 \cdots}(X(\s)) e^{i(m_2+n_2+\cdots)\s -i(m_1+n_1+ \cdots)\s}~,
\label{th-rule}
\eea
where $N$ is the number of traces in $u$ and the argument of the Dirac delta function $\dt (0)$
appearing on the right hand side is the worldsheet space direction:
\bea
\dt(0) = \lim_{\s \to \s'} \dt (\s-\s') =\lim_{\s\to \s'} {1\over 2\pi} \sum_{n\in
Z}e^{in(\s-\s')}~.
\label{delta-0}
\eea
The way one gets $N$ factors of $\dt(0)$ on the right hand side is as follows:
Each infinite-dimensional trace breaks up into a trace in the physical spacetime which appears in
the expression of $U$, and a sum over all the string modes which gives rise to a factor of
$\displaystyle{\sum_{n\in Z} 1 =2\pi \dt(0)}$. We relate the two sides of (\ref{th-rule}) by the
symbol $\sim$ to indicate that such a manipulation is understood only at a formal level. The
relation (\ref{th-rule}) implies that $u$ enjoys the same tensorial
properties in the infinite-dimensional spacetime as $U$ does in the physical spacetime (provided
$g_{ij}$ and $a^i$ have the right tensorial property, which is indeed the case as we have already
discussed).

\item
In the infinite-dimensional language the problem at hand possesses certain shift properties which
can be written as:
\bea
u^{i_1+i i_2 \cdots}_{j_1j_2\cdots} &=& u^{i_1 i_2+i \cdots}_{j_1j_2\cdots} = u^{i_1 i_2
\cdots}_{j_1-i j_2\cdots} = u^{i_1 i_2 \cdots}_{j_1j_2-i\cdots} = \cdots ~, \cr
\del_{j+l} a^{k+l}(x) &=& \del_j a^k(x)+ i(l) \delta^k_j~,
\label{shift}
\eea
where the factor of $i$ in the second term of the last equation is the imaginary number.
Given the spacetime index $i$ as in footnote \ref{index}, we have defined $(i)=m$. A shift
in the infinite-dimensional index is defined to be $i+j=\{\mu, m+n\}$\footnote{Notice that we choose the physical spacetime index corresponding to $i+j$ by the one associated with the first index (i.e. $i$) appearing in the shift. We will follow this convention in all our expressions.}. It is now obvious that the first relation
of (\ref{shift}) is a direct consequence of (\ref{th-rule}). The second relation can be obtained
from the following one:
\bea
\del_ja^k = i(j) \delta^k_j~.
\label{del-a}
\eea
The easiest way to get this is to notice that the definitions in (\ref{xga-XGdX}) imply that the
infinite dimensional model in (\ref{Linfinite}) corresponds to the string worldsheet theory only
for the linear profile $a^k(x) = (k) x^k$. Alternatively, one can directly calculate the left
hand side of (\ref{del-a}) using the third equation in (\ref{xga-XGdX}) and,
\bea
\del_i = \oint {d\s \over 2\pi}~ e^{im\s}{\dt \over \dt X^{\mu}(\s)}~, \quad {\dt X^{\mu}(\s)
\over \dt X^{\nu}(\s')} = 2\pi \dt^{\mu}_{\nu} \dt(\s-\s')~.
\label{x-X}
\eea
The result is given by $in\delta_{n,q}\delta^{\kappa}_{\nu}=i(j)\dt_j^k$.

\end{enumerate}
\section{DeWitt-Virasoro generators}
\label{s:vir-gen}

The goal of this section is to arrive at the background independent version of the quantum Virasoro generators.
We will start with the standard expressions for the classical EM tensor and write the classical Virasoro generators in the infinite-dimensional language. Then after quantizing the system we will use DeWitt's argument to define the quantum DWV generators.

The right and left moving components of the classical EM tensor are given by,
\bea
\T(\s) &=&{1\over 4} \lt( K(\s)-Z(\s)+V(\s) \rt) = \sum_{m \in Z}L_m e^{im\s}~, \cr
\tilde \T(\s) &=&  {1\over 4} \lt( K(\s)+Z(\s)+V(\s) \rt) = \sum_{m\in Z} \tilde L_m e^{-im\s}~,
\eea
respectively, where,
\bea
K(\s) &=& G^{\mu \nu}(X(\s))P_{\mu}(\s)P_{\nu}(\s) = \sum_{m\in Z} K_m e^{im\s}~, \cr
Z(\s) &=& 2 \del X^{\mu}(\s) P_{\mu}(\s) = \sum_{m\in Z} Z_m e^{im\s}~, \cr
V(\s) &=& G_{\mu \nu}(X(\s)) \del X^{\mu}(\s) \del X^{\nu}(\s) = \sum_{m\in Z} V_m e^{im\s}~.
\eea
The conjugate momentum is given by: $P_{\mu}=G_{\mu\nu}(X)\dot
X^{\nu}$. It is related to the momentum in the infinite-dimensional
language, i.e. $p_i=g_{ij}(x)\dot x^j$ according to the
same rule in (\ref{th-rule}). The classical Virasoro generators $L_m$ and $\tilde L_m$ can now be expressed
in terms of the Fourier modes $K_m$, $Z_m$ and $V_m$, which can, in turn, be expressed in the
infinite-dimensional language. The results are as follows:
\bea
4 L_{(i)} = K_{(i)}-Z_{(i)}+V_{(i)} ~, \quad
4 \tilde L_{(i)} = K_{(\bar i)}+Z_{(\bar i)}+V_{(\bar i)}~,
\label{LLtilde-classical}
\eea
where we have defined $\bar i=\{\mu, -m\}$ and,
\bea
K_{(i)} &=& g^{k l+i}(x) p_k p_l~, \quad Z_{(i)} =2 a^{k+i}(x) p_k~, \quad V_{(i)}=g_{kl}(x)
a^k(x) a^{l+i}(x)~.
\label{KZV-classical}
\eea
Notice that a Virasoro generator is a scalar, but has a string mode index $(i)=m$ which, in the
infinite-dimensional language, appears to be a shift of the spacetime index, as evident from
eqs.(\ref{KZV-classical}).

Poisson brackets of the generators in (\ref{LLtilde-classical}) should satisfy the classical
Virasoro algebra. Notice that since GCT is a canonical transformation which preserves the Poisson
brackets, it should be possible to write such brackets in a manifestly covariant manner. This has been shown in \cite{pm0912}.

We now quantize the system:
\bea
[\xh^i, \ph_j] = i \alpha' \delta^i_j ~.
\label{can-comm}
\eea
We work in the Schr$\ddot{\rm{o}}$dinger picture so that the operators
do not have explicit $\tau$ dependence.
The idea is to keep the general covariance manifest in the quantum theory. GCT of any operator independent of momenta does not suffer from any ordering ambiguity and therefore straightforward
to compute. Transformation of the momentum operator, which preserves the canonical commutation relations, is taken to be
\cite{dewitt52, dewitt57}:
\bea
\ph_i \to \ph'_i = {1\over 2} (\lam_i^{~j}(\xh) \ph_j + \ph_j \lam_i^{~j}(\xh) )~.
\label{GCT-ph}
\eea
This defines GCT of an arbitrary operator constructed out of the
phase-space variables.

Let us now introduce the position eigenbasis $|x\ra$. The orthonormality and completeness
conditions read:
\bea
\la x|x'\ra = \delta(x,x')=g^{-1/2}(x) \delta(x-x')~, \quad \int dw ~|x\ra \la x| =1~,
\label{ortho}
\eea
where $\delta (x-x')$ is the Dirac delta function, $dw = dx g^{1/2}(x)$ and $g(x) =|\det
g_{ij}(x)|$. The position space representation of the momentum
operator is given by \cite{dewitt52},
\bea
\la x|\ph_i|x'\ra = -i\alpha' \lt[\del_i + {1\over 2} \gamma_i(x) \rt] \delta (x,x')~,
\label{p-rep}
\eea
where $\gamma_i$ are the contracted Christoffel symbols\footnote{Notice that $\gamma_i$ has a
trace and therefore is divergent:
\bea
\gamma_i(x) = 2\pi \dt(0) \oint {d\s \over 2\pi}~ \Gamma_{\mu}(X(\s)) e^{im\s}~,
\eea
where the expression for $\Gamma_{\mu}(X)$ can be read out from (\ref{chr}) using the replacement
(\ref{replace}).},
\bea
\gamma_j = \gamma^i_{ji}~, \quad \gamma^i_{jk} = {1\over 2} g^{il}\lt(\del_j g_{lk} + \del_k
g_{lj} - \del_l g_{jk} \rt)~.
\label{chr}
\eea
Using $\gamma^*_i(x)=\gamma_{\bar i}(x)$ and $\del_{x^i}\delta (x,x') = - (\del_{x'^i}
+\gamma_i(x)) \delta(x,x')$ (see \cite{dewitt52}) it is straightforward
to check that the position space
representation in (\ref{p-rep}) is compatible with the following hermiticity properties:
\bea
(\xh^i)^{\dagger} = \xh^{\bar i}~, \quad (\ph_i)^{\dagger} = \ph_{\bar i}~.
\label{xp-dagger}
\eea

To construct the DWV generators we first define following \cite{omote72}:
\bea
\pih_j = \ph_j + {i\alpha' \over 2} \gamma_j(\xh)~, \quad \pih_j^{\star} = \ph_j - {i\alpha'
\over 2} \gamma_j(\xh)~.
\label{pih-def}
\eea
Using (\ref{GCT-ph}) it is easy to show that these objects are transformed by left and right
multiplications respectively under GCT:
\bea
\pih_i \to \pih'_i = \lam_i^{~j}(\xh) \pih_j~, \quad \pih^{\star}_i \to \pih'^{\star}_i =
\pih^{\star}_j \lam_i^{~j}(\xh) ~,
\label{GCT-pih}
\eea
The quantum definition of the operators in eqs.(\ref{KZV-classical}) are given by,
\bea
\hat K_{(i)} &=& \pih^{\star}_k g^{k l+i}(\xh) \pih_l~, \cr
\hat Z_{(i)} &=& \hat Z^L_{(i)} + \hat Z^R_{(i)}~, \cr
\hat V_{(i)} &=& g_{kl}(\xh) a^k(\xh) a^{l+i}(\xh)~
\label{KZV-quantum}
\eea
where,
\bea
\hat Z^L_{(i)} = \pih^{\star}_k a^{k+i}(\xh)~, \quad \hat Z^R_{(i)} = a^{k+i}(\xh) \pih_k .
\label{ZLZR}
\eea
Given the transformation properties in (\ref{GCT-pih}), it is clear that all the operators in
(\ref{KZV-quantum}) and (\ref{ZLZR}) are invariant under GCT and have the right classical limit.
The left-right symmetric combination for $\hat Z_{(i)}$ considered in (\ref{KZV-quantum}) gives the right hermiticity property for the DWV generators. These operators give covariant results in the following sense.  Consider the matrix element of an arbitrary operator constructed out
of these generators between any two scalar states. The result written in position space representation is manifestly covariant.

\section{DeWitt-Virasoro algebra}
\label{s:vir-alg}

Here we will present the algebra satisfied by the DWV generators $\hat L_{(i)}$ and $\hat{\tilde L}_{(i)}$ defined through eqs.(\ref{LLtilde-classical}, \ref{KZV-quantum}, \ref{ZLZR}). As mentioned earlier, for a generic background the current framework allows us to calculate this algebra only in the spin-zero representation. The details of the computation can be found in \cite{pm0912}. The final results, written in the infinite-dimensional language, are as follows,
\bea
\la \chi| \lt\{ \begin{array}{l}
[\hat L_{(i)}, \hat L_{(j)}] = (i-j) \alpha' \hat L_{(i+j)} + \hat A^R_{(i)(j)}~,  \cr
[\hat{\tilde L}_{(i)}, \hat{\tilde L}_{(j)}] = (i-j) \alpha' \hat{\tilde L}_{(i+j)} + \hat A^L_{(i)(j)}~, \cr
[\hat L_{(i)}, \hat{\tilde L}_{(j)}] = \hat A_{(i)(j)}
\end{array} \rt\} |\psi \ra ~,
\label{scalar-alg}
\eea
where $|\chi \ra$ and $|\psi \ra$ are two arbitrary scalar states ($\tau$-dependent). The above result is the Witt algebra (without the central charge terms) with additional operator anomaly terms given by,
\bea
\hat A^R_{(i)(j)} &=& 0~, \cr
\hat A^L_{(i)(j)} &=& 0~, \cr
\hat A_{(i)(j)} &=& \displaystyle{{\alpha'^2\over 8} \lt(\hat \pi^{\star k+i}
r_{kl}(\hat x) a^{l+\bar j}(\hat x) - a^{k+i}(\hat x) r_{kl}(\hat x) \hat \pi^{l+\bar j} \rt)}~,
\label{anomalies-id}
\eea
where $r_{ij}(x)$ is the Ricci tensor in the infinite-dimensional spacetime which, according to the general map (\ref{th-rule}), is related to the same in physical spacetime, namely $R_{\mu \nu}(X)$ in the following way,
\bea
r_{ij} (x) \sim 2\pi \dt(0) \oint {d\s\over 2\pi} ~R_{\mu \nu}(X(\s)) e^{i(m+n)\s}~.
\label{rij-Rmunu}
\eea


Notice that there are no central charge terms in (\ref{scalar-alg}). As mentioned in section \ref{s:intro}, this is due to the fact that the vacuum, which is a special spin-zero state, has not been introduced and the DWV generators have not been normal ordered. In general such a procedure is not understood to us. However, we will demonstrate this in the specific examples of flat and pp-wave backgrounds in the next section.

\section{Flat and pp-wave backgrounds as special cases}
\label{s:special}


We will discuss the flat and the pp-wave cases separately below in subsections (\ref{ss:flat}) and (\ref{ss:pp}) respectively. Before that we make some comments based on general grounds.
For both the backgrounds the EM tensor does not involve any non-trivial ordering between fields and conjugate momenta in the chosen coordinate system. 
Another way of seeing this is that in both the cases we have,
\bea
\gamma_k(x) = 0 ~, \quad g(x) = 1~.
\label{gamma-g}
\eea
This indicates that any operator calculation done using the canonical commutators without worrying about manifest general covariance should be reliable. This justifies recognizing the computation done in \cite{kazama08, pm08, pm0902} as a special case of the present analysis as will be done below. 
It also turns out as a consequence of (\ref{gamma-g}) that the algebra in (\ref{scalar-alg}) can be considered as operator equations.

\subsection{Flat background}
\label{ss:flat}

In this case the DWV generators are given, in the usual worldsheet language, as follows,
\bea
\hat L^{(0)}_m = {1\over 2} \eta_{\mu \nu} \sum_{n\in Z} \hat \Pi^{\mu}_{m-n}\hat \Pi^{\nu}_n ~,
\quad \hat{ \tilde L}^{(0)}_m = {1\over 2} \eta_{\mu \nu} \sum_{n\in Z}
\hat{\tilde \Pi}^{\mu}_{m-n} \hat{\tilde \Pi}^{\nu}_n ~,
\label{vir-gen0}
\eea
where the superscript ${(0)}$ refers to flat background and,
\bea
\hat \Pi^{\mu}_m &=& {1\over \sqrt{2}} \oint {d\s\over 2\pi} (\eta^{\mu \nu} \hat P_{\nu}(\s) -
\del \hat X^{\mu}(\s)) e^{-im\s} ~, \cr
\hat{\tilde \Pi}^{\mu}_m &=& {1\over \sqrt{2}} \oint {d\s \over 2\pi} (\eta^{\mu \nu} \hat
P_{\nu}(\s) + \del \hat X^{\mu}(\s)) e^{im\s}~,
\eea
are the usual creation-annihilation operators,
\bea
[\hat \Pi^{\mu}_m, \hat \Pi^{\nu}_n ] = \eta^{\mu \nu} \alpha' \dt_{m+n, 0} ~, \quad [\hat{\tilde \Pi}^{\mu}_m, \hat{ \tilde \Pi}^{\nu}_n ] = \eta^{\mu \nu} \alpha' \dt_{m+n, 0} ~.
\label{Pi-comm}
\eea
The DWV generators differ from the actual quantum Virasoro generators $\hat {\cal L}^{(0)}_m$ and
${\hat{\tilde {\cal L}}}^{(0)}_m$ by additive c-numbers,
\bea
\hat L^{(0)}_m = \hat {\cal L}^{(0)}_m + c_m ~, \quad \hat{\tilde L}^{(0)}_m = \hat{\tilde {\cal L}}^{(0)}_m + \tilde c_m ~,
\label{L-calL}
\eea
where,
\bea
\hat {\cal L}^{(0)}_m = {1\over 2} \eta_{\mu \nu} \sum_{n\in Z} :\hat \Pi^{\mu}_{m-n}\hat \Pi^{\nu}_n: ~,
\quad \hat{ \tilde {\cal L}}^{(0)}_m = {1\over 2} \eta_{\mu \nu} \sum_{n\in Z} :\hat{\tilde
\Pi}^{\mu}_{m-n} \hat{\tilde \Pi}^{\nu}_n: ~,
\label{vir-gen0-norm}
\eea
and $c_m = \tilde c_m$ is non-zero only when $m=0$, in which case it is a divergent constant. The normal ordering $::$ used in the above equations, which matter only for the Virasoro zero modes, is defined as the oscillator normal ordering with respect to the vacuum $|0\ra$ defined by,
\bea
\lt. \begin{array}{l} \hat \Pi^{\mu}_m \cr
\hat{\tilde \Pi}^{\mu}_m \end{array}\rt\} |0\ra = 0~, \quad \forall m\geq 0~.
\label{vac0}
\eea
It is easy to check using (\ref{Pi-comm}) that the generators in (\ref{vir-gen0}) satisfy,
\bea
[\hat L^{(0)}_m, \hat L^{(0)}_n ] &=& (m-n)\alpha' \hat L^{(0)}_{m+n}~, \cr
[\hat{\tilde L}^{(0)}_m, \hat{\tilde L}^{(0)}_n ] &=& (m-n)\alpha' \hat{\tilde L}^{(0)}_{m+n}~,
\cr
[\hat L^{(0)}_m, \hat{\tilde L}^{(0)}_n ] &=& 0~,
\label{vir-alg0}
\eea
which is simply the DWV algebra in (\ref{scalar-alg}) for flat background. However, the same method of computation applied to $[\hat {\cal L}^{(0)}_m, \hat {\cal L}^{(0)}_{-m}]$ gives a result that has operator ordering ambiguity. The result is $2m \alpha' \hat {\cal L}^{(0)}_0$ up to an additive c-number contribution that can not be calculated using this method because of the ambiguity. As indicated in \cite{polchinski}, this c-number contribution, which turns out to be the central charge term, can be found unambiguously by calculating, for example, $\la 0|\hat {\cal L}^{(0)}_m \hat {\cal L}^{(0)}_{-m} |0\ra $ with $m>0$,
\bea
[\hat{\cal L}^{(0)}_m, \hat{\cal L}^{(0)}_n ] &=& (m-n)\alpha' \hat{\cal L}^{(0)}_{m+n} + {D \over 12} (m^3 -m) \alpha'^2 \dt_{m+n, 0}~, \cr
[\hat{\tilde{\cal L}}^{(0)}_m, \hat{\tilde{\cal L}}^{(0)}_n ] &=& (m-n)\alpha' \hat{\tilde{\cal L}}^{(0)}_{m+n} + {D \over 12} (m^3 -m) \alpha'^2 \dt_{m+n, 0}~,
\cr
[\hat{\cal L}^{(0)}_m, \hat{\tilde{\cal L}}^{(0)}_n ] &=& 0~.
\label{vir-alg0-norm}
\eea

\subsection{Explaining conformal invariance for pp-wave}
\label{ss:pp}

In \cite{pm0902} we considered a restricted ansatz for an off-shell pp-wave in type IIB string theory which includes the R-R plane-wave background. The R-R flux part of the background involves the Green-Schwarz fermions on the worldsheet. We will ignore this fermionic part and consider only the bosonic part of the computation which corresponds to switching on a metric-background where the non-trivial components of the metric (in physical spacetime) are given by,
\bea
G_{+-}= 1~, \quad G_{++}=K(\vec X)~, \quad G_{IJ} = \dt_{IJ}~,
\label{G-pp}
\eea
where the vector sign refers to the transverse part with index $I$. The only non-trivial component of the Ricci-tensor is,
\bea
R_{++} \propto \vec \del^2 K~.
\label{R++}
\eea
The worldsheet theory is expected to be an exact CFT when $R_{++}$ vanishes \cite{amati88}. We call the background in (\ref{G-pp}) simply as pp-wave\footnote{The particular solution of $R_{++}=0$ given by,
\bea
K = \sum_I s_I X^I X^I ~, \quad \sum_I s_I = 0~,
\eea
is called plane-wave.}.

It was argued in \cite{pm0902} that the operator anomaly terms of the Virasoro algebra suffer from an ordering ambiguity and therefore proving conformal invariance was not completely settled. We argued below eqs.(\ref{gamma-g}) why it should be possible to consider this computation as a special case of the present background independent formulation. Here we would like to show how doing this explains the conformal invariance in the present case resolving the ambiguous situation in the previous work.

The first step is to relate the DWV generators specialized to the present background with the quantum Virasoro generators defined in \cite{pm0902}. One can check that this relation is precisely the same as that in (\ref{L-calL}),
\bea
\hat L^{pp}_m = \hat {\cal L}^{pp}_m + c_m ~, \quad \hat{\tilde L}^{pp}_m = \hat{\tilde {\cal L}}^{pp}_m + \tilde c_m ~,
\eea
where the superscript $pp$ refers to the pp-wave being considered.
According to the calculations of the present work the algebra satisfied by $\hat L^{pp}_m$ and $\hat{\tilde L}^{pp}_m$ is given by (\ref{scalar-alg}) evaluated for the pp-wave. Following the same procedure as in flat-case, which took us from eqs.(\ref{vir-alg0}) to eqs.(\ref{vir-alg0-norm}), one finds the following quantum Virasoro algebra in the present case,
\bea
[\hat{\cal L}^{pp}_m, \hat{\cal L}^{pp}_n ] &=& (m-n)\alpha' \hat{\cal L}^{pp}_{m+n} + {D \over 12} (m^3 -m) \alpha'^2 \dt_{m+n, 0} + \hat A^R_{mn}~, \cr
[\hat{\tilde{\cal L}}^{pp}_m, \hat{\tilde{\cal L}}^{pp}_n ] &=& (m-n)\alpha' \hat{\tilde{\cal L}}^{pp}_{m+n} + {D \over 12} (m^3 -m) \alpha'^2 \dt_{m+n, 0} + \hat A^L_{mn}~,
\cr
[\hat{\cal L}^{pp}_m, \hat{\tilde{\cal L}}^{pp}_n ] &=& \hat A_{mn}~,
\label{vir-alg-pp-norm}
\eea
where the operator anomaly terms $\hat A^R_{mn}= \hat A^R_{(i)(j)}$, $\hat A^L_{mn}=\hat A^L_{(i)(j)}$ and $\hat A_{mn}=\hat A_{(i)(j)}$ are given by eqs.(\ref{anomalies-id}) evaluated for the pp-wave.

We will now compare the result in (\ref{vir-alg-pp-norm}) found in this work with the one in \cite{pm0902}. The computation of \cite{pm0902} was done using the local worldsheet language. The result is precisely the local version of (\ref{vir-alg-pp-norm}) with the local operator anomaly terms given by,
\bea
&& \hat A_{there}^R(\s,\s') = \hat A_{there}^L(\s,\s')= \hat A_{there}(\s,\s') \cr
&& \propto \lt[\hat {\cal O}(\s,\s') \hat P_-(\s') \hat P_-(\s') -
\hat {\cal O}(\s',\s) \hat P_-(\s) \hat P_-(\s)\rt] \dt_{\eps}(\s-\s') \cr
&& -\lt[\hat {\cal O}(\s,\s') \del \hat X^+(\s')\del \hat X^+(\s') - \hat {\cal O}(\s',\s) \del \hat X^+(\s) \del \hat X^+(\s)\rt] \dt_{\eps}(\s-\s')~,
\label{anomalies}
\eea
where the above three anomaly terms are defined in eqs.(3.20) of \cite{pm0902} and
\bea
\hat {\cal O}(\s, \s') = \hat P_I(\s) \del_I K(\hat{\vec X}(\s')) +
\del_I K(\hat{\vec X}(\s')) \hat P_I(\s)~.
\eea
Therefore comparing (\ref{vir-alg-pp-norm}) with the corresponding result in \cite{pm0902} we are able to relate the operator anomaly terms in (\ref{vir-alg-pp-norm}) and in (\ref{anomalies}),
\bea
\hat A^R_{(i)(j)} &=& \oint {d\s\over 2\pi} {d\s'\over 2\pi} ~\hat A_{there}^R(\s,\s') e^{-im\s-in\s'} ~, \cr
\hat A^L_{(i)(j)} &=& \oint {d\s\over 2\pi} {d\s'\over 2\pi} ~\hat A_{there}^L(\s,\s') e^{im\s +in\s'}~, \cr
\hat A_{(i)(j)} &=& \oint {d\s\over 2\pi} {d\s'\over 2\pi} ~\hat A_{there}(\s,\s') e^{-im\s +in\s'}~.
\label{anomalies-rel}
\eea

It was explained in \cite{pm0902} that the expression in (\ref{anomalies}) suffers from an operator ordering ambiguity. The idea here is to resolve this ambiguity by borrowing the results found here. Therefore, using eqs.(\ref{anomalies-rel}) and (\ref{anomalies-id}) one concludes that both $\hat A_{there}^R(\s,\s')$ and $\hat A_{there}^L(\s,\s')$ should vanish. Moreover, comparing the equations in (\ref{anomalies-id}), (\ref{anomalies}) and (\ref{anomalies-rel}) one concludes that the Ricci-term obtained in the present analysis was missing earlier. As we will show below, this is an apparent discrepancy which may be resolved by showing that the relevant term vanishes, though the Ricci tensor itself does not, for the pp-wave under consideration because of certain special properties of the background\footnote{This argument, however, will rely on the scalar expectation value of the algebra in (\ref{scalar-alg}), and not operator equation.}.

The non-trivial components of the infinite-dimensional metric
corresponding to (\ref{G-pp}) are,
\bea
g_{i_+j_-} = \dt_{m+n, 0}~, \quad g_{i_+j_+} = \oint {d\s \over
  2\pi}~K(\vec X(\s)) e^{i(m+n)\s}~,\quad g_{i_{\perp}j_{\perp}} =
\dt_{IJ}\dt_{m+n, 0}~,
\label{g-pp}
\eea
where the infinite-dimensional spacetime index is divided in the
following way: $i = (i_+, i_-,  i_{\perp})$ such that,
\bea
i_+ = \{+, m\}~, \quad i_- = \{-, m\}~, \quad i_{\perp} = \{I, m\}~.
\eea
The only non-trivial components of the Ricci tensor are
$r_{i_+j_+}(\vec x)$ (the vector sign referring to the transverse
indices $i_{\perp}$) which is related to $R_{++}(\vec X)$ in (\ref{R++}) according to (\ref{rij-Rmunu}).
Let us now go back to the last equation in (\ref{scalar-alg}).
By going to the position space representation, integrating by parts
and using the shift property (\ref{shift}) one can show that the
Ricci-term is proportional to,
\bea
\int dw~ \chi^* \nabla^{k+i}(r_{kl} a^{l+\bar j}) \psi = \int dx
~\chi^* g^{k_++i k'_-} \del_{k'_-}(r_{k_+l_+} a^{l_++\bar j}) \psi~,
\eea
where on the right hand side we have evaluated the term for the pp-wave. This vanishes as the quantity inside the round brackets is independent of $x^{i_-}$.

\section{Conclusions}
\label{s:conclusions}

In this work we have explored a new approach to study two dimensional
non-linear sigma model in hamiltonian formalism. By re-writing the
problem in terms of the Fourier modes of the string we develop a
language which formally describes motion of a particle in an
infinite-dimensional curved background. A background independent
notion of the  Virasoro generators, called DeWitt-Virasoro generators,
has been defined for an arbitrary metric-background (not necessarily
on-shell) by following DeWitt's coordinate independent description of
quantum mechanics. The algebra of such generators in spin-zero
representation has been shown  to satisfy the Witt algebra with
additional operator anomaly terms that vanish for Ricci-flat
backgrounds. The same result has been shown to be true in \cite{pm1004} by
constructing tensor representations in a certain sense. 

The invariant matrix elements computed in this work involve arbitrary
spin-zero states. The field theoretic divergences are hidden in our
formal expressions as infinite-dimensional traces. However, because of
manifest general covariance one is able to carry out the necessary
derivations, at least at a formal level. Such divergences are supposed
to get cured by first introducing the right vacuum state and then
normal ordering the DWV generators. The central charge terms are also
expected to show up in the algebra when such a procedure is carried
out. Although this has not been understood in general in the present
work, it has been demonstrated for the special cases of flat and
pp-wave backgrounds. Understanding of the vacuum state and how the DWV
anomaly computed here should be related to the standard one-loop beta
function result is important for further progress.

\end{document}